# Iron valence in double-perovskite $(Ba,Sr,Ca)_2FeMoO_6$: Isovalent substitution effect

Y. Yasukawa and M. Karppinen*

*Materials and Structures Laboratory, and Department of Innovative and Engineered Materials, Interdisciplinary Graduate School, Tokyo Institute of Technology, Yokohama 226-8503, Japan*

J. Lindén

*Physics Department, Åbo Akademi, FIN-20500 Turku, Finland*

T. S. Chan and R. S. Liu

*Department of Chemistry, National Taiwan University, Taipei, Taiwan, Republic of China*

H. Yamauchi

*Materials and Structures Laboratory, and Department of Innovative and Engineered Materials, Interdisciplinary Graduate School, Tokyo Institute of Technology, Yokohama 226-8503, Japan*

In the Fe-Mo based *B*-site ordered double-perovskite, $A_2FeMoO_{6.0}$, with iron in the mixed-valence II/III state, the valence value of Fe is not precisely fixed at 2.5 but may be fine-tuned by means of applying chemical pressure at the *A*-cation site. This is shown through a systematic $^{57}$Fe Mössbauer spectroscopy study using a series of $A_2FeMoO_{6.0}$ [*A* = (Ba,Sr) or (Sr,Ca)] samples with high degree of Fe/Mo order, the same stoichiometric oxygen content and also almost the same grain size. The isomer shift values and other hyperfine parameters obtained from the Mössbauer spectra confirm that Fe remains in the mixed-valence state within the whole range of *A* constituents. However, upon increasing the average cation size at the *A* site the precise valence of Fe is found to decrease such that within the *A* = (Ba,Sr) regime the valence of Fe is closer to II, while within the *A* = (Sr,Ca) regime it is closer to the actual mixed-valence II/III state. As the valence of Fe approaches II, the difference in charges between Fe and Mo increases, and parallel with this the degree of Fe/Mo order increases. Additionally, for the less-ordered samples an increased tendency of clustering of the anti-site Fe atoms is deduced from the Mössbauer data.

* Corresponding author:
Prof. Maarit Karppinen
Materials and Structures Laboratory
Tokyo Institute of Technology
4259 Nagatsuta, Midori-ku
Yokohama 226-8503, Japan
Phone: +81-45-924-5333
Fax: +81-45-924-5365
E-mail: karppinen@msl.titech.ac.jp

## I. INTRODUCTION

The $B$-site ordered double-perovskite, $Sr_2FeMoO_{6-w}$, has been known as a ferrimagnet with a transition temperature as high as ~420 K.[1,2] Recently the phase was highlighted owing to its halfmetallic ground state and tunneling-type negative magnetoresistance (MR) behavior at temperatures up to the magnetic transition temperature.[3] The stoichiometry of the $Sr_2FeMoO_{6-w}$ phase dictates that either Fe or Mo is in a reduced valence state, *i.e.*, lower than $Fe^{III}$ or $Mo^{VI}$. As a most straightforward explanation for the halfmetallicity and the magnetic characteristics of $Sr_2FeMoO_{6-w}$, a picture in which localized $3d^5$ electrons of high-spin $Fe^{III}$ ($t_{2g}^3e_g^2$; $S$ = 5/2) and an itinerant $4d^1$ electron of $Mo^V$ ($t_{2g}^1$; $S$ = 1/2) couple antiferromagnetically (AF), was assumed.[3] However, AF superexchange interaction can not account for the metallic conductivity of $Sr_2FeMoO_{6-w}$.[4] The $^{57}$Fe Mössbauer spectra obtained for $Sr_2FeMoO_{6-w}$ samples are best interpreted by assuming a mixed-valence or valence-fluctuating state, expressed as $Fe^{II/III}$, for iron.[5,6] The valence mixing agrees with recent XANES[7,8] and field-dependent paramagnetic susceptibility[9] data. Furthermore, the mixed-valence concept is in accordance with the metallicity of the phase indicating that the mechanism behind the magnetic ordering is double-exchange interaction between the minority spins of Mo and Fe.[10] In terms of the magnitude of isomer shift a similar mixed-valence state had earlier been established by means of Mössbauer spectroscopy for the $A$-site ordered double-perovskite, $SmBaFe_2O_{5+\delta}$ ($\delta \approx 0$).[11] In the case of $SmBaFe_2O_{5+\delta}$, the $Fe^{II/III}$ state separates upon cooling into equal amounts of divalent and trivalent iron at a Verwey-type transition with $T_V \approx 230$ K,[11] while for $Sr_2FeMoO_{6-w}$ the mixed-valence state was confirmed to persist down to 5 K.[5]

In terms of fine-tuning the charge distribution "isovalent" substitution at the $A$ site is expected to provide us with interesting possibilities, as such a substitution keeps overall charge of the $B$ cations constant, but is likely to influence the distribution of charge among the two cations. Isovalent cation substitutions are sometimes discussed as chemical alternatives for application of external pressure, *i.e.*, "chemical pressure". For the case of $(R^{III}_{0.7}A^{II}_{0.3})MnO_3$ (where $R$ is a rare earth element and $A$ an alkaline earth element) magnetoresistors with a simple perovskite structure it was found that with decreasing effective ionic radius at the ($R$,$A$)-cation site, the Curie temperature ($T_C$) decreased and the magnitude of the MR effect increased drastically.[12] The primary effect of the decrease in the ionic radius at the ($R$,$A$) site was attributed to the decrease in the Mn-O-Mn bond angle, which affects the electron hopping between Mn atoms.[12] Similarly, chemical pressure has been reported to affect the value of $T_C$ as well as the low-field and high-field intergrain MR of $A_2FeMoO_{6-w}$ ($A$ = Ba, Sr and Ca) double perovskites.[13] The effect of external hydrostatic pressure on resistivity (ρ) and $T_C$ of $Ba_2FeMoO_{6-w}$ was studied as well.[14] It was found that ρ decreased and $T_C$ increased with increasing hydrostatic pressure in a parallel way to the case of chemical pressure.[14]

Since the chemical pressure through isovalent *A*-site cation substitution strongly controls the physical properties of perovskite compounds, it is interesting to investigate how it affects the valence of iron for a series of $A_2FeMoO_{6-w}$ samples. Mössbauer spectroscopy is a powerful tool for determining the electronic and spin states of Fe, and for probing the chemical environment around the Fe nuclei. Here we present the results of a detailed iron valence study for a series of $A_2FeMoO_{6-w}$ samples by means of $^{57}$Fe Mössbauer measurements and discuss the results together with data from magnetic and MR measurements.

## II. EXPERIMENTAL

A series of $A_2FeMoO_{6-w}$ samples with $A$ = Ba, $Ba_{0.8}Sr_{0.2}$, $Ba_{0.5}Sr_{0.5}$, $Ba_{0.2}Sr_{0.8}$, Sr, $Ca_{0.2}Sr_{0.8}$, $Ca_{0.8}Sr_{0.2}$ and Ca, was prepared by means of an oxygen-getter-controlled low-$O_2$-pressure encapsulation technique[15] from stoichiometric mixtures of high-purity (99.9 - 99.99 %) powders of $ACO_3$ ($A$ = Ba, Sr and Ca), $Fe_2O_3$ and $MoO_3$. The powder mixtures were calcined in air at 900 ºC for 15 hours. The calcined powders were then pelletized and sintered in an evacuated and sealed fused-quartz ampoule that contained the sample pellets together with Fe grains (99.9 % up). The Fe grains act as a getter for oxygen. The empty space inside the ampoule was filled with a fused-quartz rod in order to create a more homogeneous atmosphere in the ampoule. Each sample was fired at 1150 ºC for 50 hours. Under these conditions the oxygen partial pressure is expected to equilibrate at $2.6 \times 10^{-13}$ atm due to the redox couple of Fe/FeO.[16] After synthesis the ampoule was quenched onto a thick Cu plate immediately from the synthesis temperature. Additionally, for the Mössbauer measurements four samples with the *A*-site composition of $A$ = Sr, $Ca_{0.25}Sr_{0.75}$, $Ca_{0.5}Sr_{0.5}$ and $Ca_{0.75}Sr_{0.25}$ were synthesized using a flowing $H_2/N_2$ gas technique.[17] For these samples, the final sintering was carried out in a 5 % $H_2/N_2$ gas mixture at 1000 °C for 26 hours and then 12 hours after an intermediate grinding. The latter synthesis technique is the one commonly applied for the synthesis of $Sr_2FeMoO_{6-w}$ and related compounds. Here employment of two different techniques enables us to confirm that the obtained results for the valence of Fe are not synthesis-technique dependent.

The phase purity of the samples was confirmed by x-ray powder diffraction (XRD; MAC Science M18XHF[22]; Cu*K*α radiation). The XRD data was also used for refining the lattice parameters and cation occupancies by means of Rietveld analysis (using RIETAN program). For the oxygen-content determination, an analysis method[15] based on coulometric titration of $Fe^{II}$ and/or $Mo^{V}$ species formed upon acidic dissolution of the sample was applied. The samples were also characterized by dc magnetization (from 5 K to 400 K, and –5 T to 5 T), resistivity and MR measurements (from 5 K to 375 K, and -7 T to 7 T) using a superconductivity-quantum-interface-device magnetometer (SQUID; Quantum Design: MPMSR-5S) and a four-point probe (Quantum Design System-Model 6000), respectively.

The $^{57}$Fe Mössbauer spectra of the samples were recorded at 77 K in transmission geometry. The absorbers were made by spreading the sample materials mixed with epoxy resin evenly on an Al foil. The thickness of the sample material expressed in mass per area was approximately 20 mg/cm$^2$. A linear Doppler velocity sweeping from –11.15 to 11.15 mm/s was used. A Cyclotron Co $^{57}$Co:Rh (25mCi, January 2002) source was used for producing the 14.4 KeV Mössbauer $\gamma$ quanta. The spectra exhibiting magnetic splitting were fitted with the full Hamiltonian of combined electric and magnetic interactions. The following hyperfine parameters were included in the fit: the internal magnetic field experienced by the Fe nucleus ($B$), the chemical isomer shift relative to $\alpha$-Fe ($\delta$), the quadrupole coupling constant ($eQV_{zz}$), the resonance linewidth ($\Gamma$), and the relative intensity of the component ($I$).[5] The angle $\beta$ between $B$ and $eQV_{zz}$ was fitted when feasible, whereas the angle $\alpha$ was fixed at 0°.

## III. RESULTS AND DISCUSSION

From the x-ray diffraction patterns, zero traces of impurity phases were seen for each sample. For the $A_2FeMoO_{6-w}$ system, it has been reported that the grain size,[18,19] the degree of oxygen non-stoichiometry,[20] and the degree of Fe/Mo order at the $B$-cation site[21,22] strongly affect the magnetotransport characteristics. Thus in order to investigate "intrinsic characteristics" of the $A_2FeMoO_{6-w}$ system, the grain size, the oxygen content and the degree of Fe/Mo order need to be controlled for all the samples. The grain size of each sample was confirmed to lie in a narrow range of 2 to 6 μm from field-emission scanning-electron micrographs. From several parallel experiments of coulometric titration, the precise oxygen content of the samples of $Ba_2FeMoO_{6-w}$, $Sr_2FeMoO_{6-w}$ and $Ca_2FeMoO_{6-w}$ (synthesized by the encapsulation technique) was determined at $6\text{-}w$ = 5.98(1), 5.97(1) and 5.99(1), respectively. Besides, the $^{57}$Fe Mössbauer spectra being sensitive to the oxygen content and configuration of oxygen atoms about the Fe atoms did not reveal any presence of oxygen deficiency within the estimated detection limit of $w \approx 0.02$ for any of the samples. Thus, it is believed that all of the present samples have the same oxygen content, *i.e.*, 6-$w$ = 5.98(2). Henceforth we use "6.0" to denote the oxygen content (6-$w$) of the samples.

The concentration of "anti-site" defects, *i.e.*, Fe atoms occupying the regular Mo lattice site and *vice versa*, was estimated for the samples based on: (*i*) Rietveld refinement results of XRD data, (*ii*) measured magnitude of saturation magnetization, and (*iii*) relative amounts of different Fe species as obtained from the fittings of the $^{57}$Fe Mössbauer spectra. Rietveld refinement was carried out for the $Ba_2FeMoO_{6.0}$, $Sr_2FeMoO_{6.0}$ and $Ca_2FeMoO_{6.0}$ samples in space groups $Fm\bar{3}m$,[23] $I4/m$,[6] and $P2_1/n$,[24] respectively. According to the refinement the occupancy of Fe at the Mo site ($v$) is 0.015(1) for the $A$ = Ba sample, 0.047(1) for the $A$ = Sr sample and 0.040(1) for the $A$ = Ca sample. From the measured magnetization data the magnitude of saturation magnetization ($M_S$) was determined as the magnetization value *per* formula unit at 5 T and 5 K. The $M_S$ values obtained for the $A$ = Ba, Sr and Ca samples are 3.9(1) $\mu_B$, 3.8(1) $\mu_B$ and 3.6(1) $\mu_B$, respectively. The $M_S$ values for other samples with mixtures of Ba, Sr and Ca at the $A$ site are within 3.4~3.9 $\mu_B$. The reduction in $M_S$ as compared with the maximum value of 4$\mu_B$ is accounted for by the concentration of anti-site defects.[4,25] From the magnitude of $M_S$, the value of $v$ was estimated using a proposed relationship:[26] $M_S = (4 - 8v)$. In Fig. 1 the value of $v$, as revealed from (*i*) the XRD data through Rietveld refinement, (*ii*) the results of magnetization experiments, and (*iii*) the analysis of Mössbauer data (as given in Table 1 and discussed in detail in the following paragraphs), is plotted against the average ionic radius, $r(A^{II})$, of the $A$-site constituent for the samples synthesized by the encapsulation technique. The numbers of $v$ obtained from the three independent approaches are consistent. As shown in Fig. 1, all the samples possess high degrees of order. However, a tendency of slightly increased $v$ values is seen upon decreasing $r(A^{II})$.

Mössbauer spectra measured for the samples at 77 K were analyzed using four spectral components. A typical fitted spectrum is shown in Fig. 2. The hyperfine parameters obtained from the computer fittings of the spectra are presented in Table 1 for the $A_2FeMoO_{6.0}$ samples synthesized by the encapsulation technique. The main component (denoted M1) corresponds to $Fe^{II/III}$. The second component (denoted AS) is assigned to the anti-site Fe atoms with the valence of III. The third component (denoted AP) is assigned to the Fe atoms at anti-phase boundaries in accordance with ref. 27. The fourth component (denoted M2) is considered to be a satellite of the main component. It originates from the Fe atoms located adjacent to the anti-site Fe atoms. This is rather unusual because in perovskite oxides usually, only the influence of nearest-neighbour (oxygen) atoms gives rise to unique components in Mössbauer spectra. In other words, Fe atoms with different oxygen coordination spheres appear as well-resolved components even when spin and valence states are identical, whereas any variation at the next-nearest-neighbour (cation) site(s) merely causes line broadening of the components. However, in the present phase the nearest-neighbour oxygen configuration is identical for each Fe atom but the delicate charge balance between the next-nearest-neighbour atoms, *i.e.*, Mo or Fe, has already been shown to affect the Mössbauer spectra, *i.e.*, $Fe^{II/III}$ (component M1) or pure $Fe^{III}$ (component AS). Therefore, the M2 component is indeed believed to reflect $Fe^{II/III}$ atoms adjacent to an anti-site Fe atom. In our first study on the present phase[5] the AS component had a lower internal field value (~ 49 T) than in the present study. This was because the origin of component M2 had not been recognized yet at that time and in the analysis one of the two components was overlooked. Nevertheless, both M2 and AS indicate the presence of anti-site Fe atoms, although it is only AS that really originates from the Fe atoms at the Mo site, as M2 measures only indirectly the concentration of anti-site Fe. In the actual fittings the $\beta$ angle was free for components M2 and AP and fixed at -90° and 0° for M2 and AS, respectively. In the latter case the fixing was done due to the low intensity of component AS.

Let us now discuss the precise valence state of the majority Fe atoms corresponding to component M1. The isomer shift is the best hyperfine parameter to judge the valence state of iron. In general, the mean isomer shift values of 0.20-0.55 mm/s are expected for $Fe^{III}$ in oxides.[28] A typical value at 77 K for high-spin (HS) $Fe^{III}$ in a 5- or 6-coordinated surrounding is ~0.3 mm/s,[29] whereas for HS $Fe^{II}$ values of ~1.0 mm/s are expected.[28] Besides, from our previous study on the $Sr_2Fe(Mo_{1-x}T_x)O_6$ system with $T$ = W, Ta we learned the characteristic hyperfine parameters for different Fe species in a *B*-site ordered double-perovskite environment.[30] In this system, replacing Mo with $W^{VI}$ ($Ta^{V}$) enhances the formation of pure $Fe^{II}$ ($Fe^{III}$) at the expence of mixed-valence $Fe^{II/III}$ species. Accordingly the partially W(Ta)-substituted samples exhibited various mixtures of $Fe^{II}$ ($Fe^{III}$) and $Fe^{II/III}$. Typical isomer shift values for these in the paramagnetic state at 300 K were: 1.00 mm/s for $Fe^{II}$, 0.33 mm/s for $Fe^{III}$, and 0.55~0.70 for $Fe^{II/III}$. Based on these facts, the valence state of Fe atoms that gives rise to component M1 (with an isomer shift value in the range of 0.7 to 0.85 mm/s) is concluded to fall in between II and III; Component M1 may thus be assigned to $Fe^{II/III}$ in accordance with earlier reports.[5,6] However, the actual isomer shift value of M1 depends on the choice of the *A*-site cation. The dependence is illustrated in Fig. 3 for all the samples (synthesized by the two different techniques) where the value of the isomer shift is plotted against $r(A^{II})$. The straightforward interpretation is that the higher the isomer shift value is, the lower is the valence of the mixed-valence Fe in the sample, *i.e.*, closer to II. Thus, there is a systematic shift towards divalency upon increasing the Ba content. The isomer shift value of M1 in the $A$ = Ca sample indicates an Fe valence state close to "pure II/III". An earlier Mössbauer study indicated a lower Fe valence for Ba-based samples in accordance with the present results, though the authors suggested that $A$ = Ba leads to the actual mixed-valence state of $Fe^{II/III}$, whereas Ca- and Sr-based samples are closer to trivalency.[26] The recent results of Fe $L$-edge XANES study were interpreted such that the ground states of both $Ba_2FeMoO_6$ and $Sr_2FeMoO_6$ possess the mixed-valence $Fe^{II/III}$ configuration but with a larger $Fe^{II}$ contribution for $Ba_2FeMoO_6$ than for $Sr_2FeMoO_6$.[7]

From the Mössbauer data given in Table 1, it is seen that not only the isomer shift but also other hyperfine parameters differ rather little among the $A$ = (Sr,Ca) samples. For all the samples, the internal field values are typical of HS $Fe^{II}$ or $Fe^{III}$, except for component AP, which has an abnormally small internal field. The reasons for this are discussed in detail in ref. 27. Component AS exhibits the highest internal field values, confirming its assignment to pure $Fe^{III}$. High-spin $Fe^{III}$ with a coordination number of 5 or 6 usually has a saturation field of more than 50 T. Due to the valence mixing the internal field of component M1 is somewhat decreased, *i.e.*, it is slightly lower than 50 T. The fact that the strongest decrease in the field value is observed for the $A$ = Ba sample is in line with the judgement already made by considering the isomer shift values, *i.e.*, the valence and spin states of the mixed-valence Fe atoms are the lowest in this sample. However, the overall behaviour of the field of component M1 (see Fig. 4) requires further discussion. For the Ca-for-Sr substituted samples the isomer-shift values indicate a constant Fe valence. The local maximum in the internal field around 30 % Ca-substitution level is thus best explained by the fact that the transition from the tetragonal to cubic symmetry occurs in this region.[9] The variation in the (Sr,Ca) regime could then be purely due to the changes in the lattice symmetry. Also upon substituting Sr by Ba the field first increases, which may be accounted by the fact that the samples again become cubic. Then the strong decrease in field with a further increase in the Ba content is clearly due to the fact that the valence of Fe approaches II. Finally we note that component M2 exhibits field values intermediate between those for components, M1 and AS, as expected based on the isomer shift values for M2.

From the hyperfine parameters obtained for component M1, we may conclude that the valence balance $Fe^{II/III}$-$Mo^{V/VI}$ gradually shifts closer to that of $Fe^{II}$-$Mo^{VI}$ as the size of the $A$ cation increases. In other words, the difference between the charges of the two $B$-site cations increases as $r(A^{II})$ increases. Here it is interesting to compare this result with the data presented in Fig. 1 for the concentration of anti-site defects. For perovskite compounds, $A_2BB'O_6$, with the $B$-site occupied by two cations with equal atomic fractions, the degree of $B$-site order has been believed to depend on both the size of the cation $A$ and the charges of the cations $B$ and $B'$, such that (*i*) decreasing the size of $A$, and (*ii*) increasing the charge difference between the two $B$ cations, would promote the ordering.[31] In the present study, the concentration of anti-site defects, $v$, was found to increase with decreasing $r(A^{II})$ (Fig. 1). This is rather opposite to (*i*), but agrees with (*ii*). Therefore, the trend seen in Fig. 1 provides us with an additional confirmation for our interpretation of the Mössbauer parameters in terms of the precise valence state of iron in the $A_2FeMoO_{6.0}$ system.

Concerning the relative intensities of components M2 and AS, as M2 originates from the Fe atoms adjacent to anti-site Fe atoms, one would expect a constant intensity ratio of M2 to AS. In Fig. 5 the intensity ratio, $I$(M2)/$I$(AS), is plotted as a function of the intensity of AS for all the $A_2FeMoO_{6.0}$ samples discussed so far. In this plot we also include data for samples synthesized by the encapsulation technique under slightly different synthesis conditions in terms of temperature and/or duration. From Fig. 5, a decreasing trend for $I$(M2)/$I$(AS) is observed as the concentration of anti-site atoms increases. It is interesting that the theoretical intensity ratio of 6:1 is obtained only for samples with the smallest concentration of anti-site atoms. The most likely explanation for the decreasing trend is clustering of anti-site atoms.

Figures 6 (a) to (c) show the dependence of MR effect on the strength of applied magnetic field for the $Ba_2FeMoO_{6.0}$, $Sr_2FeMoO_{6.0}$ and $Ca_2FeMoO_{6.0}$ samples synthesized by the encapsulation technique. The magnitude of MR was defined by: MR (%) $\equiv 100(\rho_{(7T,5K)} - \rho_{(0T,5K)}) / \rho_{(0T,5K)}$, where $\rho$ is the resistivity of the sample at 5 K. All the figures clearly show tunnelling-type MR curves at low temperatures, that is, the magnitude of MR gets saturated at higher fields. The magnitude of MR at 5 K for $Ba_2FeMoO_{6.0}$ is smaller than those for $Sr_2FeMoO_{6.0}$ and $Ca_2FeMoO_{6.0}$, *i.e.*, 14.1 % for $A$ = Ba, and 17.4~17.5 % for $A$ = Sr or Ca. The difference in the magnitude of MR may be attributed to the difference in the iron valence in $Ba_2FeMoO_{6.0}$, (closer to $Fe^{II}$) and in $Sr_2FeMoO_{6.0}$ and $Ca_2FeMoO_{6.0}$ (closer to $Fe^{II/III}$), as half-metallicity in principle should depend on the degree of valence mixing. At 300 K $Ba_2FeMoO_{6.0}$ exhibits a larger MR effect than the $A$ = Sr and Ca samples. This is probably due to the fact that the CMR peak[32] seen about $T_C$ is not far. Our assumption is supported by the fact that the shape of the MR curve for $Ba_2FeMoO_{6.0}$ at 300 K lacks the typical steep low-field dependence, while it is still present for the 300-K MR curves of the $A$ = Ca and Sr samples.

## IV. CONCLUSIONS

A series of $A_2FeMoO_{6.0}$ samples was synthesized with the divalent $A$-site constituent ranging in size from 1.34 Å [= $r(Ca^{II})$] to 1.66 Å [= $r(Ba^{II})$]. In the sample synthesis, special care was taken to keep the samples as akin as possible in terms of the grain size, oxygen content and the (high) degree of Fe/Mo order. The MR characteristics and Mössbauer spectra revealed a high degree of half-metallicity and valence mixing between Fe and Mo. A systematic shift from the II/III mixed valence state towards pure II was observed for Fe with increasing Ba concentration at the *A* site, though even for the end phase with *A* = Ba a considerable degree of valence mixing was concluded. Parallel to the shift of the valence balance from $Fe^{II/III}$-$Mo^{V/VI}$ closer to that of $Fe^{II}$-$Mo^{VI}$, the degree of Fe/Mo order was found to increase. Furthermore, the present Mössbauer spectroscopic data suggested that anti-site defects at the *B*-cation site tend to cluster as their concentration increases.

## ACKNOWLEDGMENTS

The present work has been supported by a Grant-in-Aid for Scientific Research (contract No. 11305002) from the Ministry of Education, Science and Culture of Japan, and also by an International Collaborative Research Project Grant-2002 of the Materials and Structures Laboratory, Tokyo Institute of Technology. Besides, Y. Y. acknowledges support from JSPS Research Fellowship Program for Young Scientists (No. 14006635), and J. L. grants from the Scandinavia-Sasakawa foundation and the Magnus Ehrnrooth foundation.

**Table 1.** Hyperfine parameters obtained from the computer fittings of $^{57}$Fe Mössbauer spectra at 77 K for the samples synthesized by the encapsulation technique.

| | | *A* element | | | | | | | |
|---|---|---|---|---|---|---|---|---|---|
| Comp. | | Ba | $Ba_{0.8}Sr_{0.2}$ | $Ba_{0.5}Sr_{0.5}$ | $Ba_{0.2}Sr_{0.8}$ | Sr | $Sr_{0.8}Ca_{0.2}$ | $Sr_{0.2}Ca_{0.8}$ | Ca |
| | *B* (T) | 43.66(3) | 44.58(3) | 45.43(3) | 45.88(1) | 45.24(4) | 45.46(1) | 45.36(3) | 44.76(5) |
| M1 | $\delta$ (mm/s) | 0.849(4) | 0.823(3) | 0.766(4) | 0.719(1) | 0.700(6) | 0.703(1) | 0.731(4) | 0.684(5) |
| | *I* (%) | 80(4) | 80(2) | 87(3) | 84(1) | 77(4) | 82(1) | 67(2) | 80(3) |
| | e*QV*zz (mm/s) | -0.01(3) | -0.5(2) | -0.06(2) | -0.5(1) | -1.0(2) | -0.4(1) | -0.9(1) | -0.8(2) |
| | *B* (T) | 46.8(3) | 46.1(2) | 47.1(2) | 47.52(8) | 45.6(3) | 47.5(1) | 46.19(3) | 46.7(5) |
| M2 | $\delta$ (mm/s) | 0.66(1) | 0.66(2) | 0.57(3) | 0.43(1) | 0.50(4) | 0.56(1) | 0.50(1) | 0.51(3) |
| | *I* (%) | 8(2) | 7(1) | 8(1) | 7.8(5) | 11(2) | 9.3(2) | 22(2) | 8(2) |
| | e*QV*zz (mm/s) | 0.46(3) | 0.7(2) | 0.7(2) | 0.61(8) | 1.5(3) | 0.0(1) | 0.76(6) | 1.2(2) |
| | *B* (T) | 50.6(5) | 51.9(5) | 52(fix) | 52.4(2) | 51.7(2) | 52.2(3) | 51.1(2) | 52.3(4) |
| AS | $\delta$ (mm/s) | 0.32(fix) | 0.21(6) | 0.3(1) | 0.18(3) | 0.4(4) | 0.28(3) | 0.41(2) | 0.51(5) |
| | *I* (%) | 2(1) | 1.6(6) | 1.4(9) | 2.5(4) | 2.3(4) | 4.0(4) | 7.4(5) | 3(1) |
| | e*QV*zz (mm/s) | 0.2(3) | -0.1(2) | -0.1(4) | -0.3(1) | -1(2) | -0.3(1) | -0.24(6) | -0.4(2) |
| | *B* (T) | 1.9(4) | 2.6(1) | 0.8(6) | 2.4(1) | 1.2(3) | 3.4(2) | 2.6(4) | 1.6(3) |
| AP | $\delta$ (mm/s) | 0.38(5) | 0.39(2) | 0.18(5) | 0.37(2) | 0.34(4) | 0.23(6) | 0.14(3) | 0.38(4) |
| | *I* (%) | 10(1) | 11.2(5) | 3.7(6) | 5.7(3) | 9(1) | 5.1(3) | 3.6(2) | 9.1(8) |
| | e*QV*zz (mm/s) | -0.8(2) | -0.48(9) | 1.0(2) | 0.3(1) | -0.8(1) | 1.1(2) | 1.1(2) | 1.2(1) |

## Figure captions

**Fig. 1.** Concentration of anti-site defects ($v$) *versus* ionic radius of cation $A^{II}$ [$r(A^{II})$]. The value of $v$ is estimated based on the measured $M_S$ value, the result of Rietveld refinement of XRD data, and the observed relative intensity of component AS in the 77-K $^{57}$Fe Mössbauer spectrum.

**Fig. 2.** Typical fitted 77-K $^{57}$Fe Mössbauer spectrum for an $A_2FeMoO_{6.0}$ sample. The spectral components used in the fit are displayed above the data points. Starting from the topmost they are: M2, AS, AP, and M1.

**Fig. 3.** Isomer shift value of component M1 *versus* ionic radius of cation $A^{II}$ [$r(A^{II})$].

**Fig. 4.** Internal field of component M1 *versus* ionic radius of cation $A^{II}$ [$r(A^{II})$].

**Fig. 5.** Intensity ratio of component M2 to AS, $I$(M2)/$I$(AS), as obtained from the fitted 77-K Mössbauer spectra for the $A_2FeMoO_{6.0}$ samples.

**Fig. 6.** The field dependence of MR effect for the samples: (a) $Ba_2FeMoO_{6.0}$, (b) $Sr_2FeMoO_{6.0}$, and (c) $Ca_2FeMoO_{6.0}$, synthesized by the encapsulation technique.

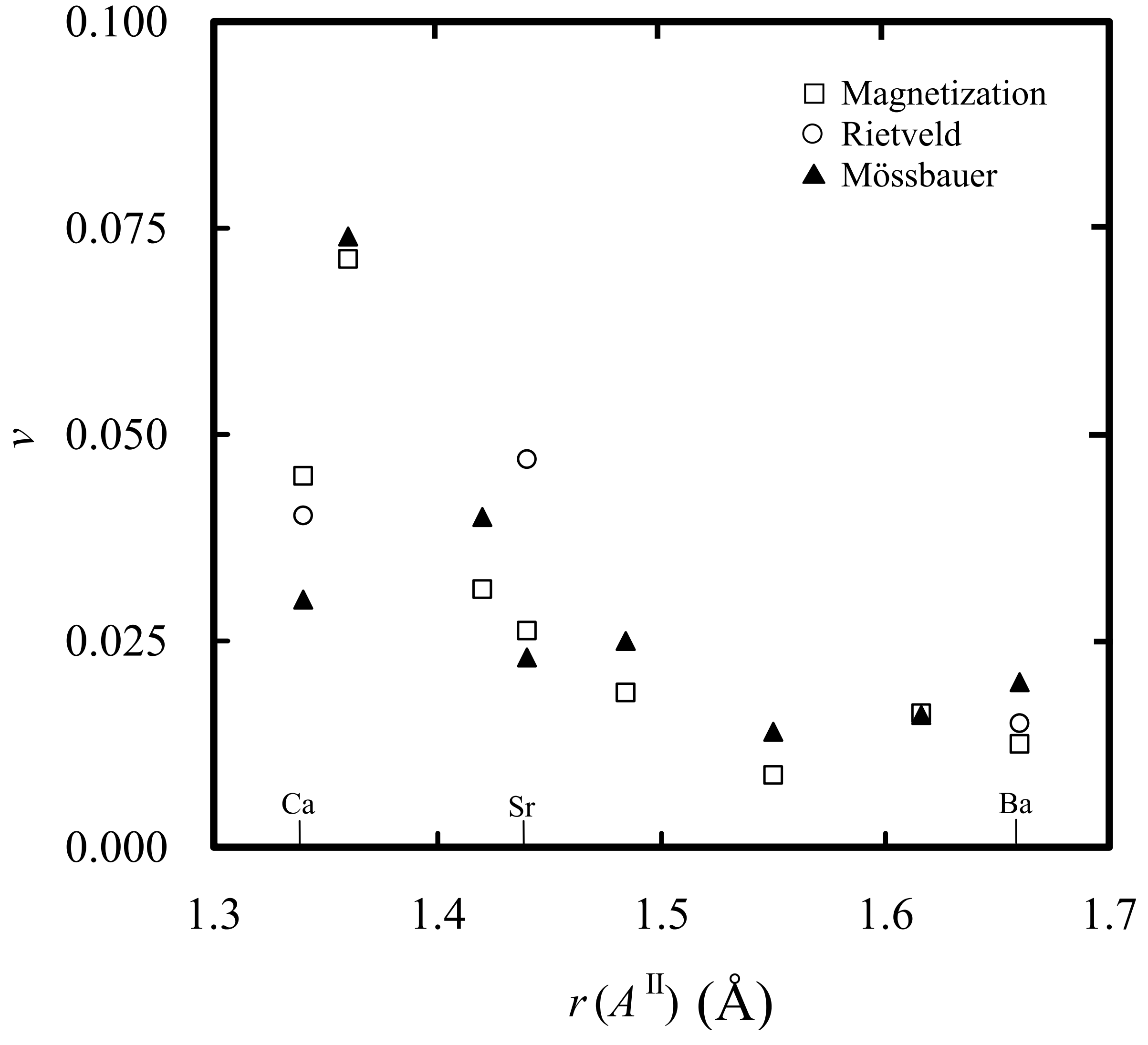

Yasukawa *et al*. (*Phys. Rev. B*) : Fig. 1.

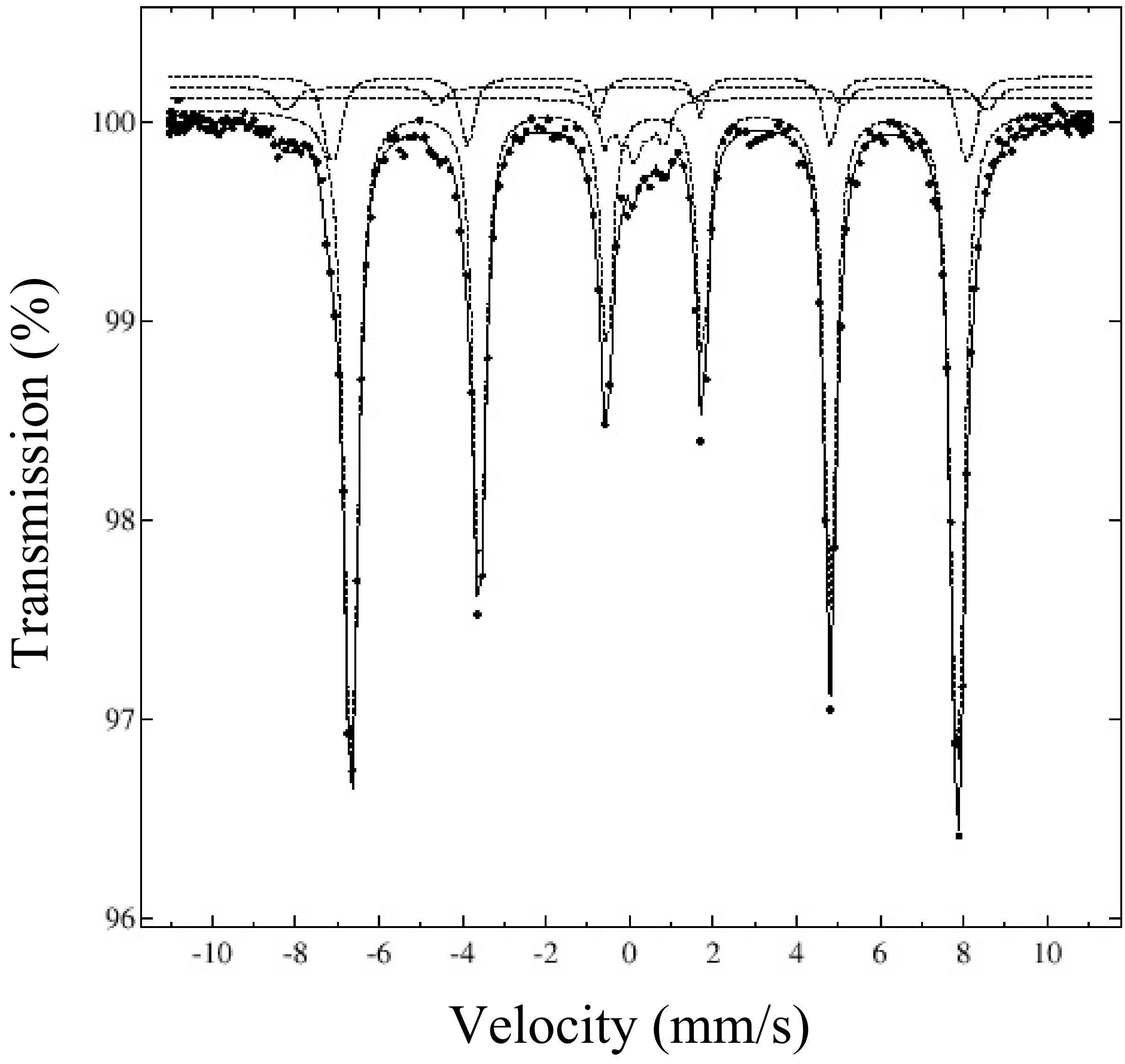

Yasukawa *et al*. (*Phys. Rev. B*) : Fig. 2.

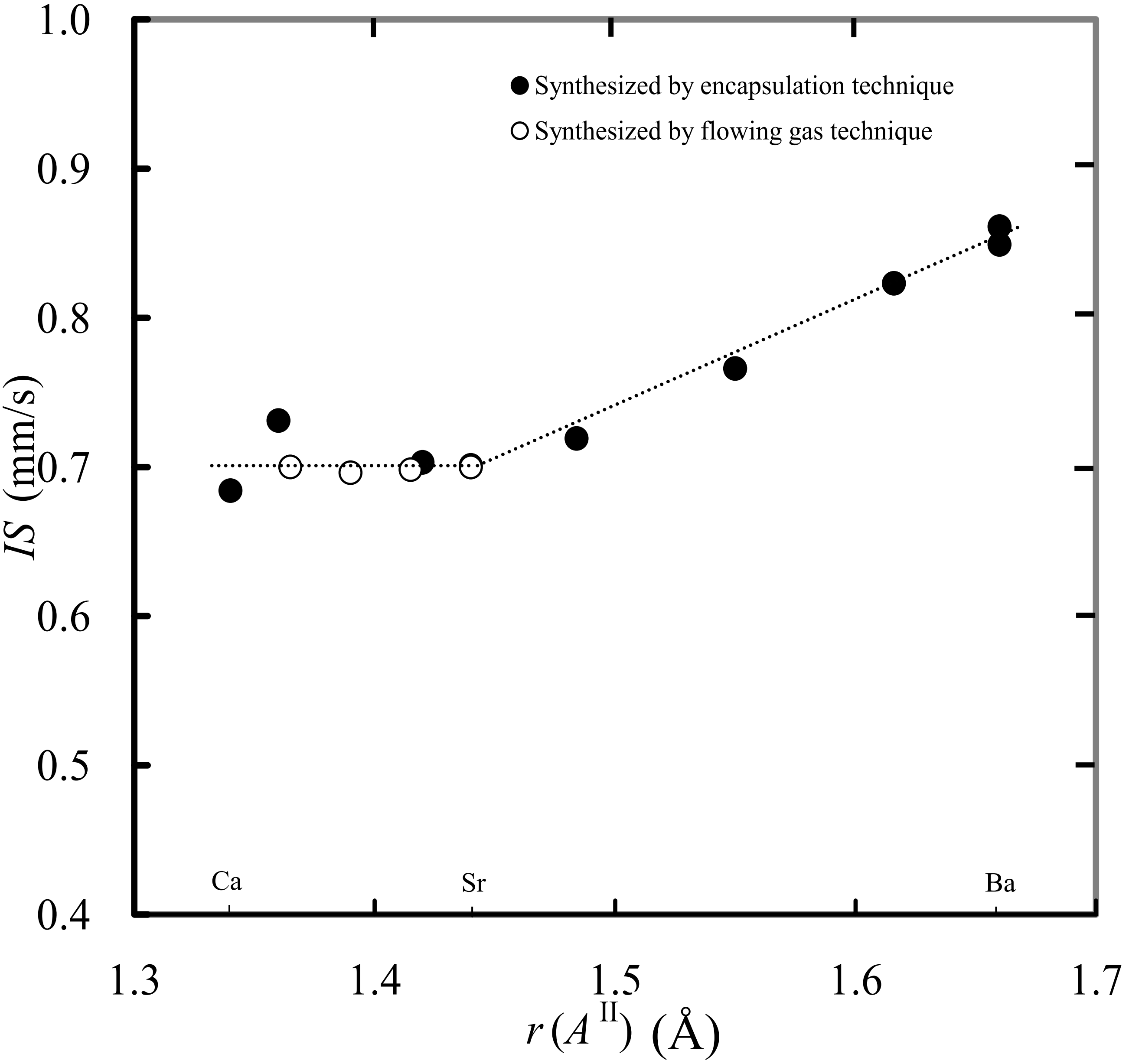

Yasukawa *et al*. (*Phys. Rev. B*) : Fig. 3.

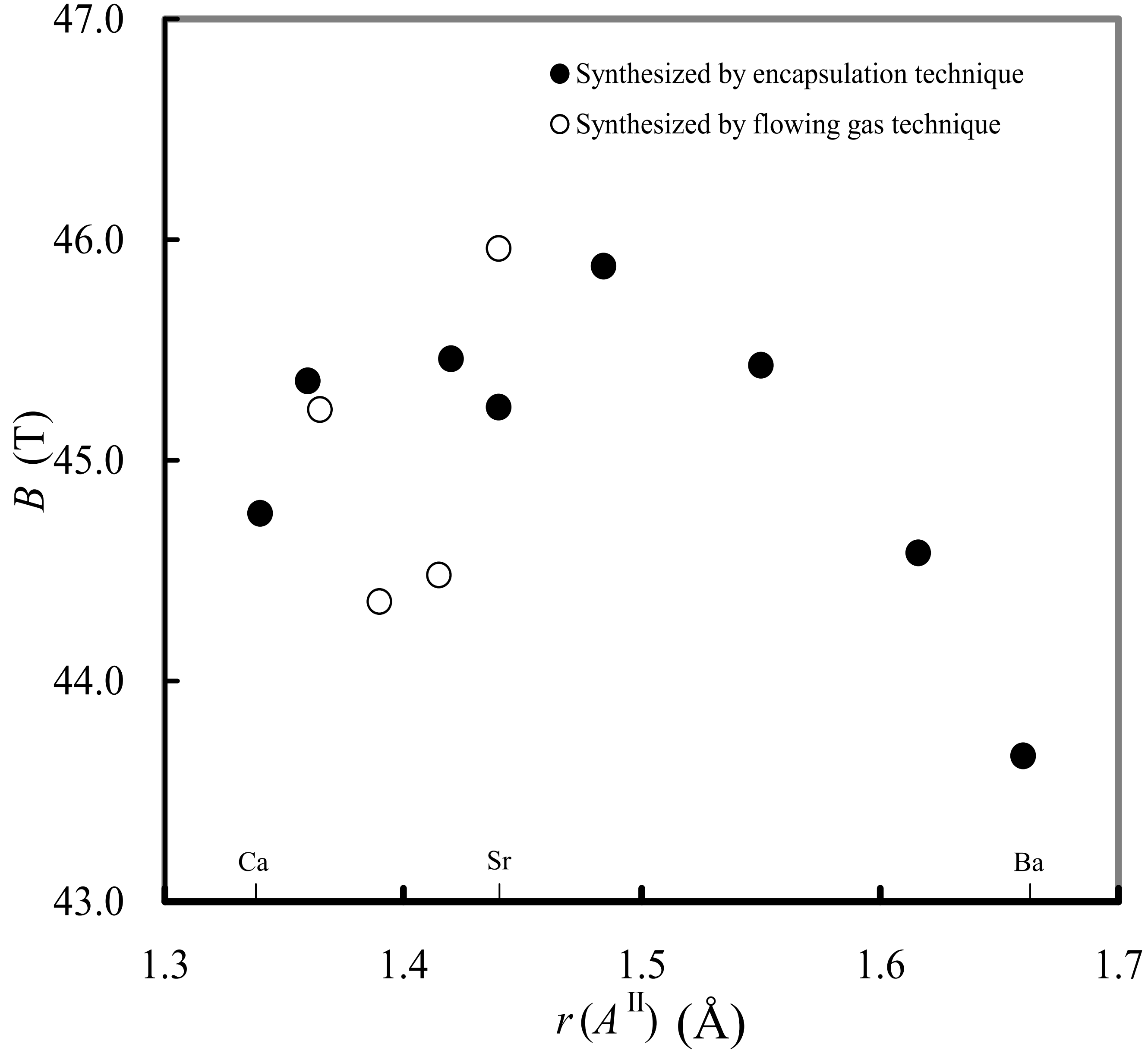

Yasukawa *et al*. (*Phys. Rev. B*) : Fig. 4.

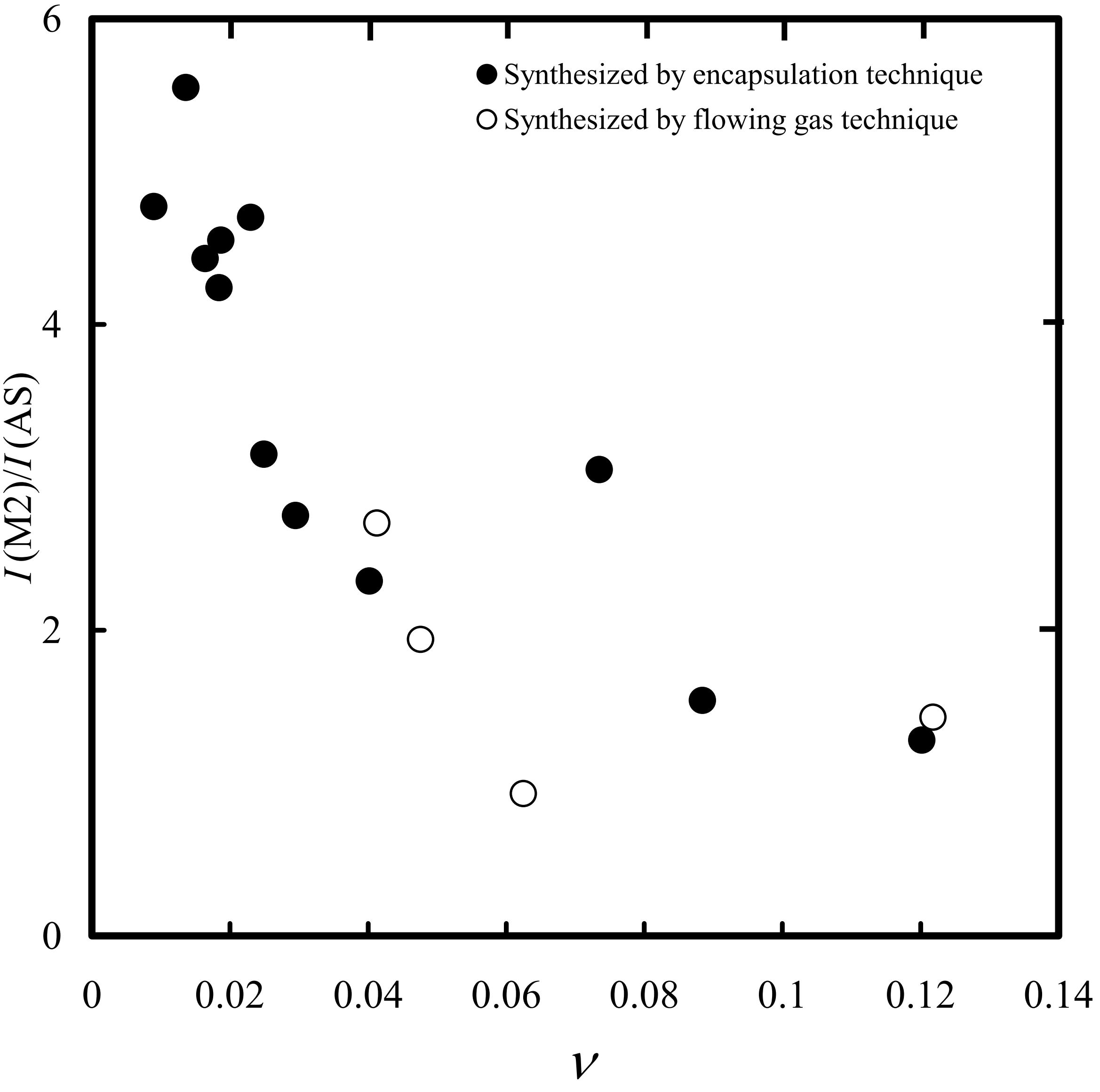

Yasukawa *et al*. (*Phys. Rev. B*) : Fig. 5.

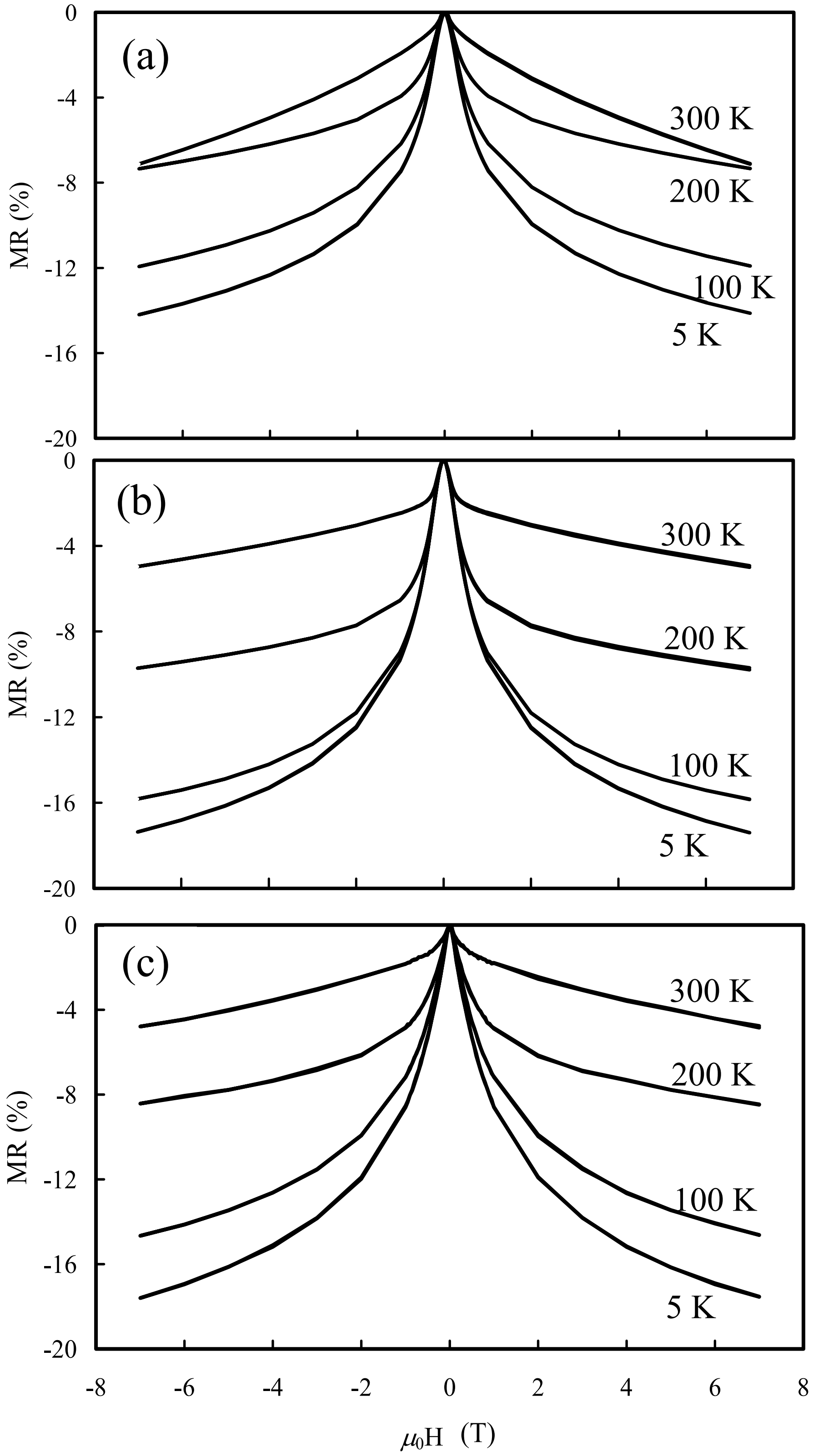

Yasukawa *et al*. (*Phys. Rev. B*) : Fig. 6.